# The aftermath of Big Deal cancellations and their impact on interlibrary loans

Marc-André Simard[1], Jason Priem[2], and Heather Piwowar[2]
[1]École de bibliothéconomie et des sciences de l'information, Université de Montréal, Québec, Canada.
[2]Our Research(https://ourresearch.org/)
Contact: marc-andre.simard.1@umontreal.ca

## Abstract

A "Big Deal" is a bundle of journals that is offered to libraries by publishers as a "one-price, one size fits all package" (Frazier, 2001). There have been several accounts of Big Deals cancellations by academic libraries in the scientific literature. This paper presents the finding of a literature review aimed at documenting the aftermath of Big Deal cancellation in University Libraries, particularly their impacts on interlibrary loan services. We find that many academic libraries have successfully cancelled their Big Deals, realizing budget savings while limiting negative effects on library users. In particular, existing literature reveals that cancellations have a surprisingly small effect on interlibrary loan requests. The reviewed studies further highlight the importance of access to proper usage data and inclusion of community members of the community (staff, faculty members, students, etc.) in the decision-making process.

## Introduction

Originally coined by Frazier (2001), a "Big Deal" is a bundle of journals that is offered to libraries by publishers as a "one-price, one size fits all package". There have been several accounts of Big Deals cancellations by academic libraries in the scientific literature (e.g. Calvert et al., 2014; Pedersen et al., 2014; Nabe and Fowler, 2012; 2015). With the increasing prices of journal subscriptions (Kyrilidou, 2012; Haigh, 2016) and possible incoming budget cuts caused by the COVID-19 pandemic, several academic libraries will not be able to continue to afford their subscriptions to Big Deals. This paper presents the finding of a literature review aimed at documenting the aftermath of Big Deal cancellation in university libraries and their impacts on interlibrary loan (ILL) services. Table 1 presents a summary of those findings.

## Successful Big Deal cancellations

In 2012, the Mississippi State University Library faced a budget shortfall of $500,000 at the same time as two of their Big Deals packages (Springer and Wiley) were up for renewal (Jones et al., 2013). Using download statistics from recent years, they established an average price-per-use for each title. They cancelled both Big Deals and purchased around 200 individual titles, saving approximately $400,000 and losing 2800 journals including all social science journals. The library retained perpetual access to some titles, but only for older material. They authors have not reported any information on ILL. The authors also regret using only usage statistics to determine the final mix of titles for the library. They wish they would have had the time to involve the faculties and other librarians in the process. They received mixed feedback: librarians have been "disheartened" by the loss of so much content, while the teaching faculty has been offered the option to swap titles. Students did not express any opinions on the unsubscriptions.

Another well-documented case of Big Deals cancellation is from the Southern Illinois University Carbondale Library. Their experience was fully documented by Nabe and Fowler (2012; 2015) over a period of five years. In 2009–2010, the library had to unsubscribe from 1 100 Springer titles (10,000 download/year), 597 Wiley titles (11,252 downloads/year) and 242 Elsevier titles (19,452 downloads/years). They did retain perpetual access to all unsubscribed titles. After one year, they observed a small increase in ILL requests. Only 27% of the unsubscribed Wiley titles had at least one request, and 9% had more than one for a total of 71 ILL requests. Wiley titles ILL demand represented 0.9% of its prior use. Among Elsevier titles, 38% had at least one ILL request, and 20% had more than one for a total of 46 requests. This represented 0.3% of its prior use. The community reaction to the cancellation was minimal. They received a small number of complaints; most being attributed to confusion or a lack of information. After one year, they estimated savings of $300,000 annually.

 Five years after cancellation, the picture had changed somewhat, with cancelled titles generating more ILL requests. Among lost Wiley titles, for instance, 32% now had at least one request, with the highest rate being 9.4 requests per year.  The authors reported a total of 1,118 ILL requests for the cancelled Wiley titles over the five years. This same list of titles received 11,254 downloads in the year prior to the cancellation. So, the authors estimate that after five years, cancelled titles have generated about 10% as many ILL requests as compared to pre-cancellation annual downloads. This is in contrast to the numbers after one year, when the authors estimated that ILL requests were only 0.3% of the pre-cancellation usage count.

Using data from three universities in the University of California System, Calvert et al. (2014) studied the effect of cancellations on their ILL services. Between 2010 and 2011, each of the three libraries unsubscribed from 100 to 800 titles. In the next twelve to eighteen months after the cancellations, they observed a small increase in the number of ILL transactions, with requests for cancelled titles accounting for 1-4% of total transactions.  Given these two facts, we can estimate that requests for cancelled titles directly increased ILL traffic by no more than 4%. Moreover, only 2 to 4% of the cancelled titles received any ILL requests.

Between 2008 and 2011, the Health Sciences Library at the University of Tennessee Health Science Center had to cancel 30% of their journal collection (Fought, 2014). They decided to use a slightly different method: instead of cancelling, they used a pay-per-view system based on tokens. They compared the costs of this system against subscriptions and ILL costs pre and post cancellation of their Wiley subscriptions. Pay-per-view articles were paid with tokens that would make an article available for 24 hours. Token price varied from $12.25 to $32.25 depending on how many they ordered. They cancelled each journal with a cost-per-use of $30.00 per article, for a total of $147.999 annually or the equivalent of 12,081 tokens (at $12.25 per). In 2010, they would have saved $110,416 if they had used tokens instead of subscriptions. In 2012, their pay-per-view costs were $156,322.25 compared to the $1,265,866 they would have paid if they had subscribed to the 764 journals. The total number of access rose from 3,068 to 12,761 from 2010 to 2012. Their analyses showed no abuse of the pay-per-view service. Considering the copyright fees of ILL requests, the authors estimate that the pay-per-view token system also allowed them to save money in ILL fees.

In 2009, the Iowa State University (ISU) allocated $2.65 million (USD) of its yearly material budget to three Big Deals with Springer, Wiley and Elsevier (Pedersen et al., 2014). They compared the average price interlibrary loan (from $13.94 to $17.50) to the cost-per-use of their journals in order to determine if they were getting their money's worth. They respectively cancelled 84% and 63% of their Springer and Wiley titles. With their savings, the library reinvested their funds on the addition of new subscriptions identified as in high demand by the community. They also retained perpetual access to their Springer journals. These cancellations had "little to no discernible" effect on ISU Library's ILL services: ILL borrowing dropped by 10% in 2012, 12% in 2013 and 19% in 2014 despite cancelling 1500 Springer journals in 2011. In 2012 and 2013, the library paid $793.60 in royalties to Springer for access to cancelled journal articles compared to their original subscription costs of $1,823.00 per year.

In 2012, the University of Memphis Library had to cancel 277 titles as reported by Knowlton et al. (2015). Out of those 277 cancelled titles, they retained perpetual access to 187 titles with the remaining 90 titles being already covered through aggregated databases. Their ILL statistics showed that the cancellations had very little effect upon ILL usage: only 6 out of 187 cancelled titles were requested (3%) which account for 0.2% of all ILL requests in the first half of 2013 (16 out of 3,845 requests).

In 2014, the University of New Mexico Health Sciences Center Library had to deal with a shortfall of $250,000 in their budget (Nash & McElfresh, 2016). Using cost-per-use data from 2013–2014, they cancelled titles with a cost per use greater than $20 and fewer than 100 yearly uses. They also issued an online survey asking respondents to rank the importance of journals to their work. They ultimately cancelled 18 journals, including 8 from the survey answers. In 2015, they received 43 ILL requests from 18 cancelled titles, which accounted for 1.4% of all ILL requests. Two articles were requested more than five times. The ILL request rate rose 137% in 2015 which also coincided with a migration towards a new integrated system with a "Request Interlibrary Loan" button. Controlling for this increase, we can estimate that ILL for cancelled titles rose by 1.4% x 137% = 1.9%. Librarians did not receive any direct request for subscriptions to the cancelled titles.

Another method shown in Raymond (2017) and Knowlton et al. (2015) is cancelling individual serial subscriptions based on the title availability in aggregated full-text databases. Since 2005, the Santa Clara University Library has saved up to $22,750 annually after cancelling a total of 75 subscriptions. They did not receive a single complaint by students of faculties since they have started cancelling those titles. They did not receive a single ILL request for those titles either.

In 2017, Kansas State University faced a $285,000 budget cut (Hoeve, 2018). After cancelling 490 journals or 67% of their collection, they saved $128,608.39. They have yet to assess the impact of those cancellations on their ILL service.

The West Virginia University Library cancelled three Big Deals: they retained 79 out of their 2,143 Springer titles, 113 out of their 1,275 Wiley titles and 541 out of their 1,412 Elsevier titles (Arnold & Mays, 2019). In 2018–2019, they saved $200,000, $441,450, and $800,000 from Springer, Wiley and Elsevier respectively. The expected increase in ILL requests has not materialized, making ILL much more cost efficient than the former subscription costs. They received a total of 180 requests for Springer titles but they only paid for one ($39.95), 310 requests for Wiley and paid for 27 articles ($853.90 total costs, with an average of $31.63), and 3,604 requests for Elsevier titles but only paid for 198 of them ($4,400.70 or an average of $22.23). To this date, they received very few complaints from students and faculties about the cancellation.

In Sweden, the Bibsam consortium cancelled their Big Deals with Elsevier in 2018 for a total of 2,362 titles lost (Olsson et al., 2020). They retained perpetual access to all of their titles in exchange for a fee of 0.06 euros per download for titles. Nine months after the cancellation, there was no increase in ILL or article deliveries. Deliveries have gone up after the subscription of large research institutions to the service, driving the average monthly fee from 26,000 euros to 40,000. After surveying researchers, 54% of them claimed that the cancellation had harmed their work, while 38% of them said that it had no impact. Nearly half of the respondents had a negative view of the cancellation, while 37% had not. When searching for articles, 81% of researchers faced an Elsevier paywall, 42% of researchers found at least one missing article online, and 42% gave up their search at least once. Ultimately, this cancellation led to a new Read and Publish agreement with Elsevier in 2019 for a slightly cheaper price than their previous subscription. They project savings of 1.7 million euros in 2022 in subscription and APC fees.

## Libraries who have kept their Big Deals

Other studies have decided to keep their subscriptions after assessing them. At the University of South Alabama Biomedical Library between 2010 and 2012, articles from Big Deals showed an average cost-per-use of $6.04 (varying from $2.11 to $9.42) against an average of $15.35 for ILL and $37.72 for their pay-per-view service (Lemley & Li, 2015). They concluded that while Big Deals were cost effective for them, they limit their flexibility to purchase other resources.

Using data from the Hofstra University Library, Glasser (2013) found that between 14 and 59% titles from their four Big Deal packages were not used by its community, with 39% to 86% of their title being labeled as "low use." Between 5 and 29% of their titles were considered "high use." The average cost-per-use per title varied from $4.59 to $9.44. The author concluded that the library is getting a good deal for 3 of their 4 packages: they estimated that keeping only the "high use" title would cost between $27,000 to $882,000 more than keeping the Big Deals "as is."

Table 1. Reviewed cases of Big Deal cancellations and their effect on ILL requests.

| Study | Library | Time since cancellation | Journals cancelled (n) | Savings (approx.) | Pre-cancellation assessment method | ILL request change (percent +/-) |
|---|---|---|---|---|---|---|
| Jones et al. (2012) | Mississippi State University Library | Retroactive (2008-2010) | 2,800 | $400k | Cost-per-use | - |
| Nabe and Fowler (2012; 2015) | Southern Illinois University Carbondale Library | 2 years; 5 years | 1,939 | $320k | - | +186% from 2010 to 2014 |
| Calvert et al. (2014) | University of California System Library (n = 3) | 12-18 months | 1,084 | $305k | Cost-per-use; alignment with curricula; eliminating duplicate | 1-4% |
| Fought (2014) | Health Sciences Library at the University of Tennessee Health Science Center | 1 year | 700+ | $110k (2011); $1.11 million (2012) | Cost-per-use | decreased (exact numbers not reported) |
| Pederson et al. (2014) | Iowa State University | 5 years | 1,598 (Springer) | $300k | Cost-per-use/ILL data | -19% |
| Knowlton et al. (2015) | University of Memphis Library | 1 year | 277 | - | ILL data; Availability in full-text databases | -28% in 2013; 0.2% of ILL made to cancelled titles |
| Nash & McElfresh (2016) | University of New Mexico Health Sciences Center Library | 1 year | 18 | - | Cost-per-use; user survey | 1.9%, controlling for new ILL systems |
| Raymond (2017) | Santa Clara University Library | 10 years | 75 | $23k | Availability in full-text databases | 0% |
| Hoeve (2018) | Kansas State University | - | 490 | $129k | Qualitative assessment; Cost-per-use; Alignment with curricula | - |

| Arnold & Mays (2019) | West Virginia University Library | 2 years (Wiley); 1 year (Springer & Elsevier) | 4,097 | $1.44 million total | Cost-per-use; References made by faculty members; Qualitative assessment | Percent change not reported. Very small cost increase ($4440.65 total) |
|---|---|---|---|---|---|---|
| Olsson et al. (2020) | Bibsam consortium (Sweden) | 9 months | 2,362 | $2 million (including APCs) | - | No increase in ILL requests after 9 months; +54% (estimated) after new university subscriptions to ILL services |

*Notes.* APC = Article processing charges. Monetary amounts are in USD per year (not inflation-adjusted) unless otherwise noted. "Time since cancellation" is how long ago the cancellation happened, at the time of the study.

# Conclusion

There have been several cases of academic libraries who successfully cancelled their Big Deals while maximizing savings and limiting the effects or its patrons. Some libraries concluded that it was best for them to keep their Big Deals after assessing their impact on their collection and on their community. However, these examples have all underlined the importance of having access to proper usage data and to include members of the community (staff, faculty members, students, etc.) in the decision process. Cancelling a Big Deal is a tough decision to make, but fortunately, there are more and more useful resources (tools, softwares, scientific literature and reports, etc.) that can assist librarians in going through that process.

# Acknowledgement

The authors of this paper would like to acknowledge *Our Research* for the funding and support.

## Author order
The second and third authors contributed equally, and their order was determined by coin flip.